\documentclass{article}

\usepackage{arxiv}

\usepackage[utf8]{inputenc} 
\usepackage[T1]{fontenc}    
\usepackage{hyperref}       
\usepackage{url}            
\usepackage{booktabs}       
\usepackage{amsfonts}       
\usepackage{nicefrac}       
\usepackage{microtype}      
\usepackage{lipsum}		
\usepackage{graphicx}
\usepackage{doi}

\title{Dynamic Virtual Power Plant: A New Concept for Grid Integration of Renewable Energy Sources}

\author{ \href{https://orcid.org/0000-0000-0000-0000}{\includegraphics[scale=0.06]{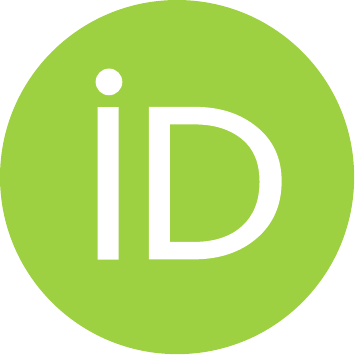}\hspace{1mm}B. Marinescu}\thanks{Corresponding author Bogdan.Marinescu@ec-nantes.fr} \\
	Ecole Centrale Nantes-LS2N, France\\
	\And
	\href{https://orcid.org/0000-0000-0000-0000}{\includegraphics[scale=0.06]{orcid.pdf}\hspace{1mm}O. Gomis-Bellmunt} \\
	CITCEA-UPC Barcelona, Spain\\
	 \And
	\href{https://orcid.org/0000-0000-0000-0000}{\includegraphics[scale=0.06]{orcid.pdf}\hspace{1mm}F. D\"orfler} \\
	ETH Zurich, Switzerland\\
	\And
	\href{https://orcid.org/0000-0000-0000-0000}{\includegraphics[scale=0.06]{orcid.pdf}\hspace{1mm}H. Schulte} \\
	HTW-Berlin, Germany\\
	\And
	\href{https://orcid.org/0000-0000-0000-0000}{\includegraphics[scale=0.06]{orcid.pdf}\hspace{1mm}L. Sigrist} \\
	UPC-IIT, Madrid, Spain\\
}

\renewcommand{\shorttitle}{\textit{arXiv} Template}

\hypersetup{
pdftitle={A template for the arxiv style},
pdfsubject={q-bio.NC, q-bio.QM},
pdfauthor={David S.~Hippocampus, Elias D.~Striatum},
pdfkeywords={Renewables, Grid integration, Grid ancillary services},
}

\begin{document}
\maketitle

\begin{abstract}
	The notion of Virtual Power Plant (VPP) has been used many times in last years in power systems and for several reasons. As a general trend, the behavior of a classic synchronous generator is to be emulated for a class of conventional grid components like, e.g., renewable generators or/and power electronic units. Most of the times production of these units is of interest, as it is the case for the new AGC scheme of Spain which, from this point of view, looks like a VPP. However, \textit{dynamic aspects} are of high importance, especially for increasing the actual rate of penetration of Renewable Energy Sources (RES). Indeed, to go above the actual rate of RES penetration, one should deal with \textit{full} participation of RES to grid services. This means not only to get some positive impact on grid voltage and frequency dynamics but to bring concepts which allows one to integrate RES to existing secondary regulation schemes on the same level as the classic synchronous generators. For that, we propose here a new concept called Dynamic VPP (DVPP) which fully integrates the dynamic aspects at all levels: locally (for each RES generator), globally (for grid ancillary services and interaction with other neighbor elements of the grid) and economically (for internal optimal dispatch and participation to electricity markets). A DVPP is a set of RES along with a set of control and operation procedures. This means methodologies for: choosing the participating RES, optimal and continuous operation as a whole (especially in case of loss of natural resources - e.g., wind, sun - on a part of the DVPP), regulation (in the dynamic sense) to ensure local objectives for each generator, participation to ancillary services of the DVPP as a unit and to diminish negative effects of interaction with neighbor dynamics elements of the power system, integration in both actual power systems scenarios (with mixed classic and power electronics based generation) and future ones with high degree of RES penetration. Concrete structures of DVPP as well as ways to address the other control and economical aspects will be shown.
This new DVPP concept is now under development in the H2020 POSYTYF project (https://posytyf-h2020.eu/).
\end{abstract}


\section{Introduction} 

\begin{figure}
    \centering
    \includegraphics[scale=0.4]{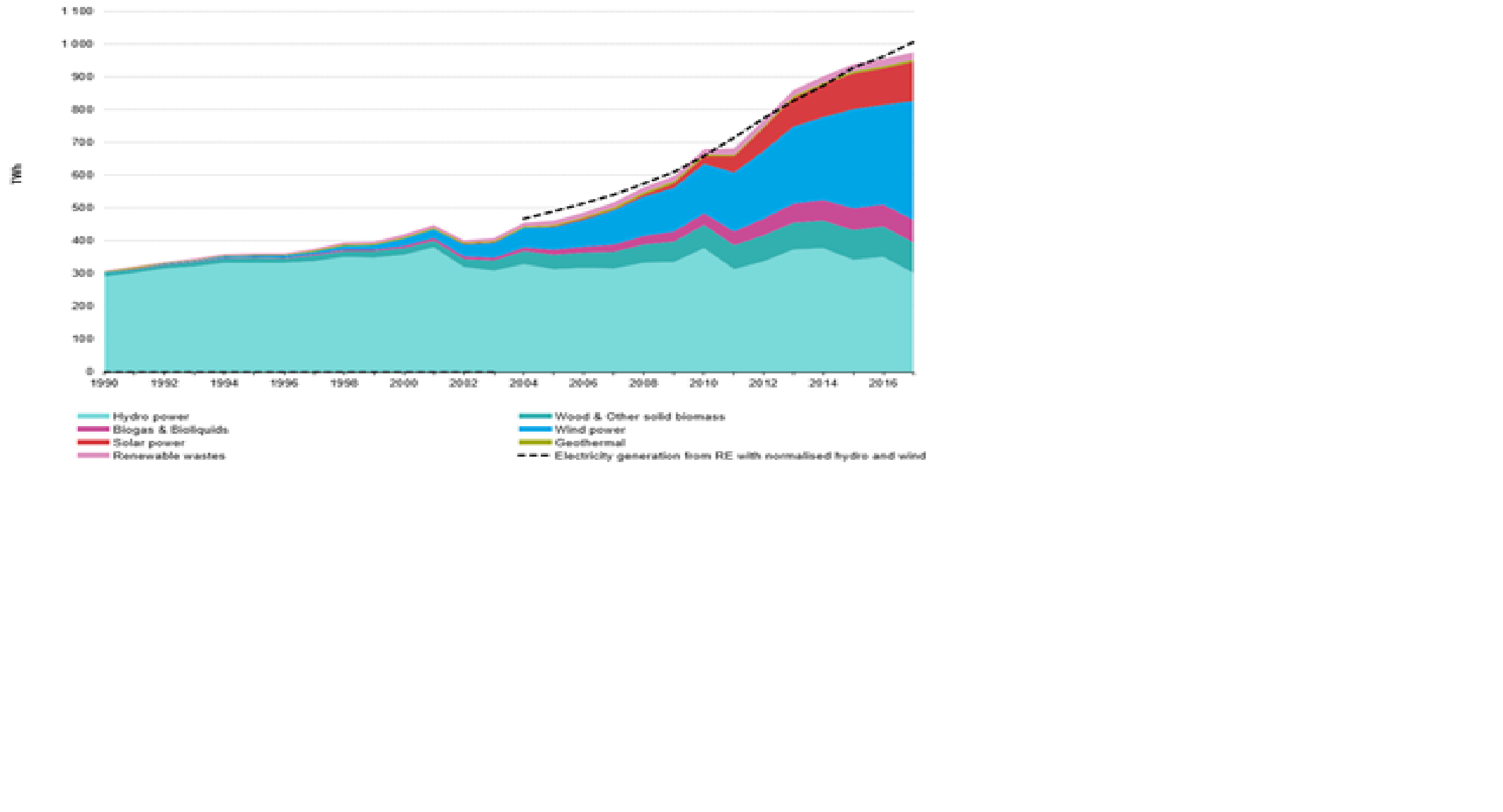}
    \caption{RES production in Europe (source: Eurostat http://mission-innovation.net/about/)} 
    \label{fig:RES_Europe}   
\end{figure}

Renewable Energy Sources (RES) are key means of global energy transformation. The volume of RES was increasing last decades in all power systems. Fig. \ref{fig:RES_Europe}  shows that, in Europe, the RES share nearly doubled from 2005 to 2015. By using more renewables to meet its energy needs, the European Union (EU) lowers its dependence on imported fossil fuels and makes its energy production more sustainable, in line with Energy Union priority. 

The EU is on track to meet its target of at least 32\% share of renewable energy in 2030 and two thirds in 2050 set in the new revised Renewables energy directive (2018/2001) and the EU SET Plan \cite{EUroadmap}. Although impressive progress has been achieved as a result of the ambition and vision to meet climate targets, more effort will be needed to meet long-term decarbonisation objectives. By 2050, renewable energy could be the largest source of energy supply, representing two-thirds of the energy mix. This requires an increase in renewables’ share of about 1.2\% per year, a seven-fold acceleration compared to recent years.
For these objectives, one must lead the development of the next generation of renewable technologies, but also integrate the energy produced from RES into the energy system in an efficient and cost-effective manner.

\begin{figure}
    \centering
    \includegraphics[scale=0.4]{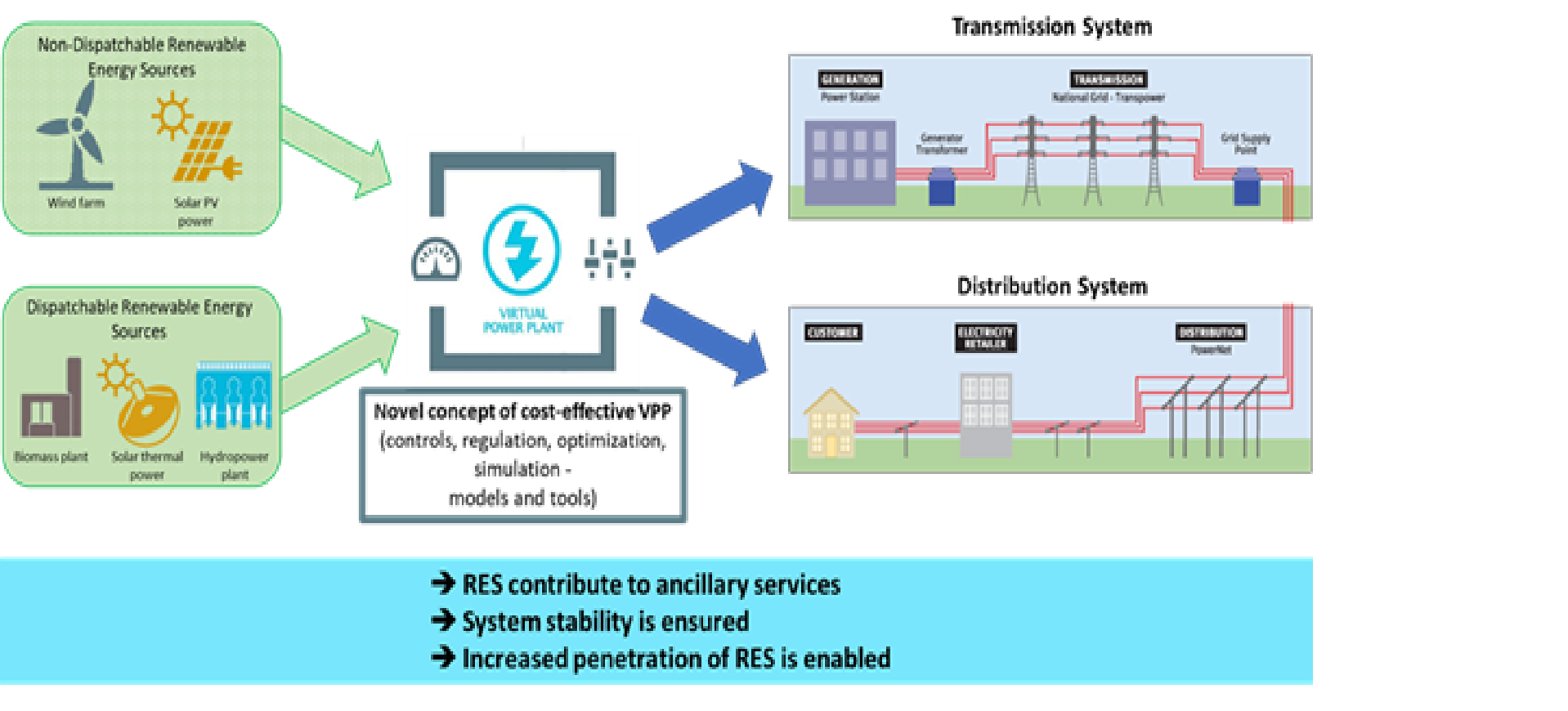}
    \caption{DVPP Concept a)} 
    \label{fig:DVPP_Concept1}   
\end{figure}

However, a high share of variable renewable generation will pose new challenges for the operation of power systems. A key question is whether there will be sufficient power system flexibility (i.e., ability to constantly keep power supply and demand in balance, responding to (quick and large) changes in either. Flexibility can be provided by generators (fossil but also dispatchable RES), consumers, storage systems, networks or even system operation rules) to deal effectively with the increased variability in generation expected.

A power system is composed of transmission and distribution grids. Transmission grid operates at Very High Voltage (VHV) levels to transmit energy over long distances, with minimal losses. Distribution grid is the end segment of the grid and provides most part of the customers connected in low voltage (some industrial customers may be connected in VHV directly to the transmission level). Importantly, they are operated by distinct actors: transmissions grids are run by Transmission System Operators (TSO) while the distribution ones by Distribution System Operators (DSO).  RES can be connected on both grids: large farms (and systematically all offshore wind parks) are directly connected to transmission grids, while lower level RES generators are more broadly spread on the distribution grid.

The stability of the electrical networks is a quality of their regulation by which the moderately disturbed situations return progressively to a state of equilibrium. The regulation (servo-control process acting on a dynamic system) aims to keep the entire network frequency and voltage magnitudes close to their set points.
Stability is a major concern of power system operation: the system has to reach a stable operation point after any disturbance (short-circuit, line or generator trip, etc). This constraint impacts:
\begin{itemize}
	\item The device – generator level: it is required that the physical variables of the machine (voltage and currents) remain within security limits so that the material is not damaged in case of disturbance and could continue to run
	\item The system – grid level: it is required that the overall balance generation/load is respected for the whole interconnected system. This means that physical variables of the grid (voltages, currents and frequency) remain within security limits, so called system or ancillary services. Therefore a global – system – view of all interconnected power system is required (not only on one or some generators).
\end{itemize}

RES are i) intermittent and causing fluctuations on all time scales which need to be balanced by dispatchable generation, ii) spatio-temporally correlated, iii) geographically dispersed (on both transmission and distribution parts), and iv) interfaced with power electronics thereby lacking the physical robustness (inertia) of rotational generation. Therefore, power transmission is required to pursue stable production with RES. However, the grid’s capacity should be high enough to transmit this power under dynamic stability constraints (dynamic limit of power lines). This limit is systematically lower than the physical thermal limit of the transmission lines wires and it is difficult to assess and ensure by regulation. One alternative to RES power transmission is storage, but this is a high cost solution. In addition, large-scale use of electrochemical batteries may have a significant environmental impact. 
Altogether, the stability issues related to RES limit their use. Indeed, in many power systems around the globe (such as Ireland, Australia, or small islanded grids) ensuring system stability is the main bottleneck to further integrate sustainable RES. A solution to overcome this, is to increase the share of so-called \textit{dispatchable} RES, i.e., the ones which have a natural energy storage capacity (solar thermal or hydropower plants, for example).

RES grid integration faces major limitations when high penetration is expected (more than 50\%). They are mainly due to several aspects of stability assessment and sureness:
\begin{itemize}
	\item dynamic stability margin (safe power transmission) should be ensured in case of meteorological hazards. One should thus prevent any risk of generalized black-out.
	\item RES should systematically participate to ancillary services: this is not the case today since RES are not majority at the overall scale of the interconnected systems. In case of increase of the share of RES this situation should change. RES connection on both transmission and distribution sides is a supplementary difficulty since, traditionally, ancillary services are provided at the transmission level.
	\item Frequency stability - in terms of electric synchrony, i.e., keeping the electric frequency in a tight band around a target (50Hz in Europe) as a necessary and sufficient condition of stability - raises a question since RES are systematically connected to the grid via power electronics and have naturally low or zero inertia. Massive integration of RES poses not only a question of stability but also the question of revisiting from a theoretical point of view such notion and conditions of stability.
\end{itemize}

We present here a new concept called Dynamic Virtual Power Plant (DVPP) to tackle the aforementioned challenges to large-scale implementation of the RES. A DVPP is, as generally shown in Fig. \ref{fig:DVPP_Concept1}, a set of different nature (dispatchable and non-dispatchable) RES. They are well chosen in order to ensure safe and optimal grid insertion and operation by offering their combined flexibility (ramping up and down at short notice for frequency control), internally balancing heir fluctuations, and selling their aggregate generation output in the wholesale market. 

The notion of Virtual Power Plant (VPP) was already used in literature but mostly for \textit{static} aspects. Indeed, the existing work on VPP deal with a set of RES generators but limited to economic dispatch (e.g., \cite{VPP3}) or RES integration in electricity markets \cite{VPP4}, \cite{VPP5}, \cite{VPP6}. The fastest dynamics studied in a VPP concern the secondary frequency-power control \cite{VPP_Spain1}, \cite{VPP_Spain2}. 

Power electronics used systematically to connect RES generators to the grid brings fast dynamics which are essential to be taken into account. In the DVPP presented here, \textit{all dynamics} of the DVPP and neighbour AC grid are taken into account. This is mandatory for a full integration of the DVPP to existing control (primary and secondary) schemes in order to allow full participation of the DVPP to grid ancillary services. Our DVPP concept integrates all aspects: static (load-flow), optimal (perimeter definition for short/medium and long-term run) and dynamic (control for local - machine- and global - grid - objectives).

The paper is structured as follows: in Section \ref{sectionSpecsObj} the specifications and objectives of the DVPP are explained. In Section \ref{sectionComponents} the components of the DVPP are given. Section \ref{sectionStatic} deals with the static aspects of the DVPP and Section \ref{sectionDynamic} with the dynamic ones. In Section \ref{sectionApproaches} are discussed the approaches used to reach the DVPP objectives while Section \ref{sectionConclusions} is devoted to conclusions.

\section{Specifications and objectives}\label{sectionSpecsObj}

The main objective of the DVPP is to integrate a portfolio of RES (including dispatchable and non-dispatchable), to provide the power system with signals able to participate to ancillary services and allowing flexibility. 

Starting from limiting facts of today’s situation, specific scientific and technical objectives are deduced. \\

\textit{\textbf{Fact 1:}} the increase of the dispatchable RES share allows the grid connection of a higher non-dispatchable RE share and thus the significant increase of the overall installed RES. 

Implementation of the general DVPP idea above requires first coordination between the 2 kinds of RES and along with conventional generation. Thus, the aggregated DVPP may present itself to the grid as a single dispatchable and fully controllable source. Resources of the DVPP may thus be optimized.

First, they may participate also to ancillary services.\\

\textit{\textbf{Fact 2:}} to allow a high penetration level of RES, such sources should participate in ancillary services.

To that end, the controls implemented for such services should be revised. The prevalent multi-layer structure should evolve to become compliant with the DVPP notion above. Also, the time scales of the controls should be revisited to deal with the fast dynamics of most of the RES (interfaced with power electronics).\\

\textit{\textbf{Objective 1:}} define structure and controls for DVPP to fully participate to grid ancillary services.\\

Stability definitions (as well as assessment) in actual power systems rely on the notion of synchrony. The latter is ensured conventionally by the large rotational inertia of synchronous generators. In case of high penetration of RES, a stable grid frequency should be ensured also by other means \cite{Milano2018}. Power electronics could be operated in grid forming mode, i.e., like a voltage source. Stability analysis and control under current saturations (“hard” limitations) is a difficult task to be addressed. The DVPP would have a main role in settling such new synchrony. This is a matter of control. But also analysis of such new notion of stability is to be done in a new manner (to be defined). Indeed, fast dynamics due to high penetration of RE and power electronics no longer allow for usual hypothesis in stability studies like the ones which led to classic classification of frequency, voltage and small-signal stability.\\

\textit{\textbf{Fact 3:}} RES grid insertion cannot rely on synchrony and "\ grid frequency"\ as inputs/hypothesis.  \\

\textit{\textbf{Objective 2:}} new means (methodologies) for analysis and assessment of stability should be introduced for the control of the DVPP.\\

RES are geographically distributed. The individual injections are small in comparison to those of classic thermal generators, but the sum of all these injections is important at the scale of the overall system. To include them in a DVPP concept, \textit{methodologies to aggregate specifications of control} are needed. 

On the one hand, the control actions should extend also to voltage and transient dynamics. In presence of massive power electronics penetration, this leads to multi-scale (fast/slow) problems. Classic assumptions of decoupling and non-interaction no longer apply. Indeed, fast operation of converters (due to operation switching or other external commutations due to, e.g., rapid setpoint changes communicated from a higher system operation layer such as pricing signals) would lead to currently rarely encountered and possibly even unforeseen types of interactions. One such class consists on coupling modes studied in \cite{Kouki2018}, \cite{Arioua2016}, \cite{Munt2017}. These are electric interactions between geographic distant devices which are different from the classic inter-area modes put into evidence between large inertia thermal synchronous generators. Thus, such interactions should be taken into account not only among RES of the DVPP but also with other external elements like, e.g., converters of HVDC links in the neighbour of the DVPP.
On the other hand, the amount of RES connected at distribution level being important, the DVPP concept should include those generators as well. This means that the aggregation logic should cover both distribution and transmission levels of the grid.  Such an aggregation is addressed in \cite{Bel2018}. It should be extended here to include the analysis and control points raised above. Moreover, to physically apply the controls to the RES generators, a disaggregation methodology of the DVPP controls should be proposed. Notice that there is no available general methodology for \textit{disaggregation of dynamic control} and this should be developed. Also, \textit{resilience} is an important item to be integrated: if a lot of local controllers are designed so that the global DVPP response is as desired, the closed-loop response should still be satisfactory if any single one of them fails.\\

\textit{\textbf{Fact 4:}} distribution connected RES units should participate to new ancillary services plan and DVPP.\\

\textit{\textbf{Objective 3:}} define a way of aggregating DVPP objectives and actions compliant with the split of the grid into transmission and distribution levels.\\

To capture the time-scales and phenomena mentioned above, new models are needed. Indeed, hypothesis of separation between voltage and frequency phenomena is no longer valid in the new context. Models should be revisited both at simulation and control levels/purposes.

\textit{\textbf{Fact 5:}} hypothesis of separation between voltage and frequency phenomena should be revisited.\\

\textit{\textbf{Objective 4:}} propose models adequate to the multi-scale and coupling dynamics of the new grids.\\
 
\textit{\textbf{Objective 5:}} define the perimeter of DVPP (to ensure economic efficiency).

This should be done both for long-term and real-time. DVPP resources portfolio should be optimized in function of availability of DVPP sources (related to meteorological conditions and to maintenance/failure constraints), grid conditions and market prices. This leads to a real-time redispatch tool which must assess both economic and security (N-1 stability) issues.

Long-term optimality of the solutions should be analyzed, especially against solutions using electrochemical storage.\\

\textit{\textbf{Objective 6:}} Prove that the proposed solution is competitive compared with solutions combining variable RES with electrochemical storage.\\

DVPP is a new concept which brings together generation and grid aspects. Moreover, as RES can be connected both to transmission and distribution grids, the DVPP perimeter may contain both types of grid. This raises thus also regulatory questions.  Proposals should be made at this level to facilitate DVPP insertion.

\textit{\textbf{Objective 7:}} Provide business cases and regulatory solutions to allow DVPP development.\\

The DVPP concept should be flexible enough to allow implementation in several stages which could progressively be followed by the TSOs, DSOs and generators:
\begin{itemize}
	\item Applicable today in the actual regulatory framework and structure 
\item Near future scenario in which RES penetration will overpass the threshold to invalidate the classic hypotheses of dynamic behaviour of interconnected power systems: higher frequency variation, separation between V/f dynamics (Fact 5) and classification of stability in transient, small-signal and voltage. Objectives 4 and 5 should be treated in this new context in a different way.
\end{itemize}

\textit{\textbf{Objective 8: }} Implementation in two stages for TSOs, DSOs and generators.\\

\begin{figure}
    \centering
    \includegraphics[scale=1.3]{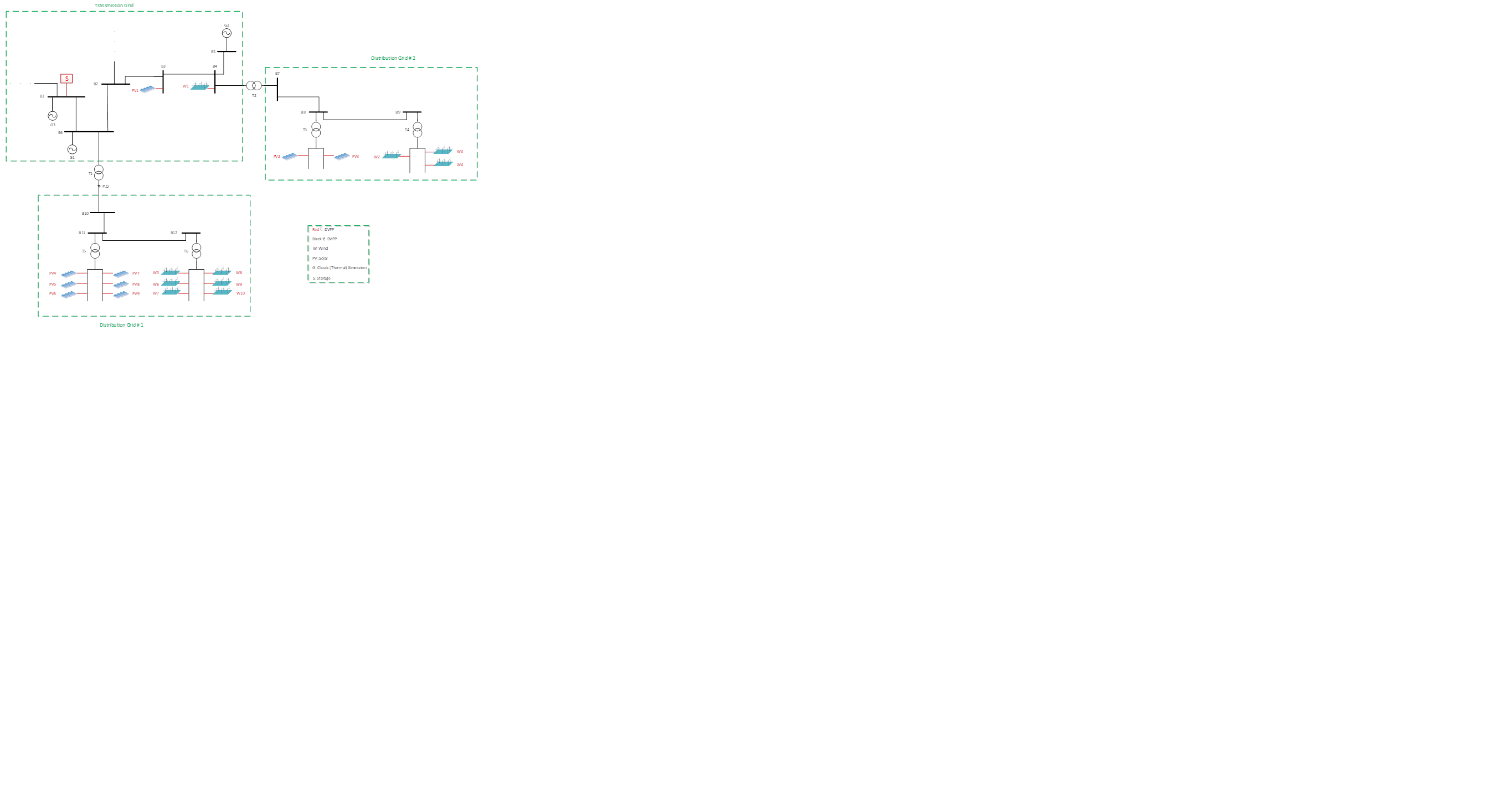}
    \caption{DVPP concept b)}
    \label{fig:DVPP_concept}
\end{figure}

A generic example of a DVPP developed to meet all above objectives is given in Fig. \ref{fig:DVPP_concept}. Indeed, it contains several kinds of RES (different technologies for different natural ressources, of dispatchable and non-dispatchable type), geographically spread on both transmission and distribution levels. Several points of grid connection may exist. DVPP generators are thus not necessarily close one to each other. Conversely, some RES generators can be close to other dynamic elements of the grid (other generators, FACTS, HVDC, ...) that do not belong to the DVPP.

\section{Components}\label{sectionComponents}

The proposed DVPP concept is based on the following types of units: 
\begin{itemize}
    \item PV: solar photovoltaic power plants (large scale)
    \item ST: solar thermal power plants, including thermal energy storage in molten salts.
    \item W: offshore or onshore wind power plants
    \item HYD: hydropower power plants
    \item PS-HPP: pumped-storage hydropower with bidirectional operation
    \item BIO: biomass power plants
    \item GEO: geothermal power plants
    \item Conventional thermal units already existing in the system can also be considered, and they can be integrated in the DVPP and coordinated with the renewable units:  CF-TPS coal-fired thermal power station, CC-TPS combined-cycle thermal power station and N-TPS nuclear thermal power station 
    \item Additional units like batteries, hydrogen electrolyzer, flexible loads, etc... can be potentially added to the concept 
\end{itemize}

The previous units can be classified in terms of dispatchability as follows:
\begin{itemize}
 \item  The primary energy availability permanently constraints the power output capability. PV 
 \item  The primary energy availability constraints the power output capability, but the power can exceed the threshold temporarily (short time-seconds).  W
 \item  The primary energy availability influences the power output capability. However, the power output can be increased by means of a secondary (inherent storage) energy source. ST 
 \item  The primary energy availability is sufficient to not constrain the output power. However, in some cases the time response will be slow (see Table \ref{tab:tab1}). BIO HYD CF-TPS CC-TPS N-TPS
 \item The primary energy availability does not constraint the power output capability and it is possible to reverse the power plant to produce primary energy from the surplus of electricity in the network (bidirectional capability).  PS-HPP
\end{itemize}

Table \ref{tab:tab1} shows the response time, the inherent storage time and the generation technology  employed by the different components discussed. Response time is understood as the time elapsed between the acknowledgement of a new power reference and its successful tracking. Inherent storage time is the total amount of time in which an electricity generation technology can provide electricity at full capacity by means of its inherent energy storage. The generation technologies considered are PE: power electronics, SG: synchronous generator and IG: induction generator. These aspects determine the role that each technology may have within the electric power system. PV and wind present fast response times (from milliseconds to a few seconds), whereas the other technologies are much slower as they are solely based on synchronous generators. However, the inherent storage time of PV and wind is zero, whereas the other technologies offer this characteristic, from hours to months (conventional plants).

\begin{table}[h]
\begin{tabular}{|l|lll|}
\hline
& Response  &	Inherent 	& Generation  \\
& time & storage time & technology \\
\hline
PV     & 100 ms - 5 s  & 0 &  PE  \\
ST     & 15 min – 4 h  & 0 - 24 hours  &  SG \\
W      & 0.5 ms - 1 s  & 0  &  SG/IG+PE  \\
HYD    &  2 - 5 min & 4h - 16h   & SG    \\
BIO    & 10 min – 6 h  & weeks  & SG   \\
CF-TPS & 80 min - 8 h  & months  & SG   \\
CC-TPS &  5 min – 3 h & months  & SG   \\
N-TPS  & $\approx$ 24 h & months  & SG   \\
PS-HPP & 2 - 5 min & 4h - 16h &  SG   \\
GEO    & 30 s – 2 min & $\infty$ & SG  \\
\hline
\end{tabular}\label{tab:tab1}
\end{table}

\section{Static aspects}\label{sectionStatic}

\subsection{Topologies and scenarios}
\label{sub:scenarios}

The DVPP concept introduced here is flexible in the sense that covers several power systems situations:

\begin{itemize}

	\item \textit{continental and island power systems}: it can insert a set of RES in an interconnected power system or in an isolated island. In the first case, it will participate to existing control schemes for the large thermal plants. In the second case, it will directly ensure voltage and frequency services.
	
	\item \textit{transmission and distribution grids}: RES can be connected on both transmission and distribution sides as shown in Fig. \ref{fig:DVPP_concept}. The new DVPP concept should allow participation of RES generators from both sides. This implies coordination of the control actions through the border between the two grids. This coordination is intended at both administrative (share of the data/measures and control actions) and technical (different voltage levels and different structure (radial for distribution versus meshed for the transmission) of the grids) levels. Data availability is important, especially for second level controls (to ensure ancillary services) and for coordination of control actions in general (as DVPP actuators are geographically distant). It is supposed that voltage and frequency measures from both transmission and distribution sides will be available at a common control point called \textit{DVPP dispatching} in the sequel. Controls will be computed in the DVPP dispatching and sent back to the DVPP actuators. The abovementionned measures can be classic or PMU.

	\item \textit{several grid connection points:} insertion of the DVPP in the rest of the system may be via several connections points as in Fig. \ref{fig:DVPP_concept}. Moreover, the DVPP may have RES generators in several distribution grids.
	
	\item \textit{imbricated structure}: RES generators included into the DVPP are not chosen from geographical or topological considerations. As a consequence, components of a DVPP are not necessarily neighbours. Moreover, some neighbour generators may not participate to the DVPP (devices in black in Fig. \ref{fig:DVPP_concept}). They should be considered as disturbances/dynamic interactions in synthesis of the DVPP controls.
	
\end{itemize}

\begin{figure}
\centering
\includegraphics[scale=0.2]{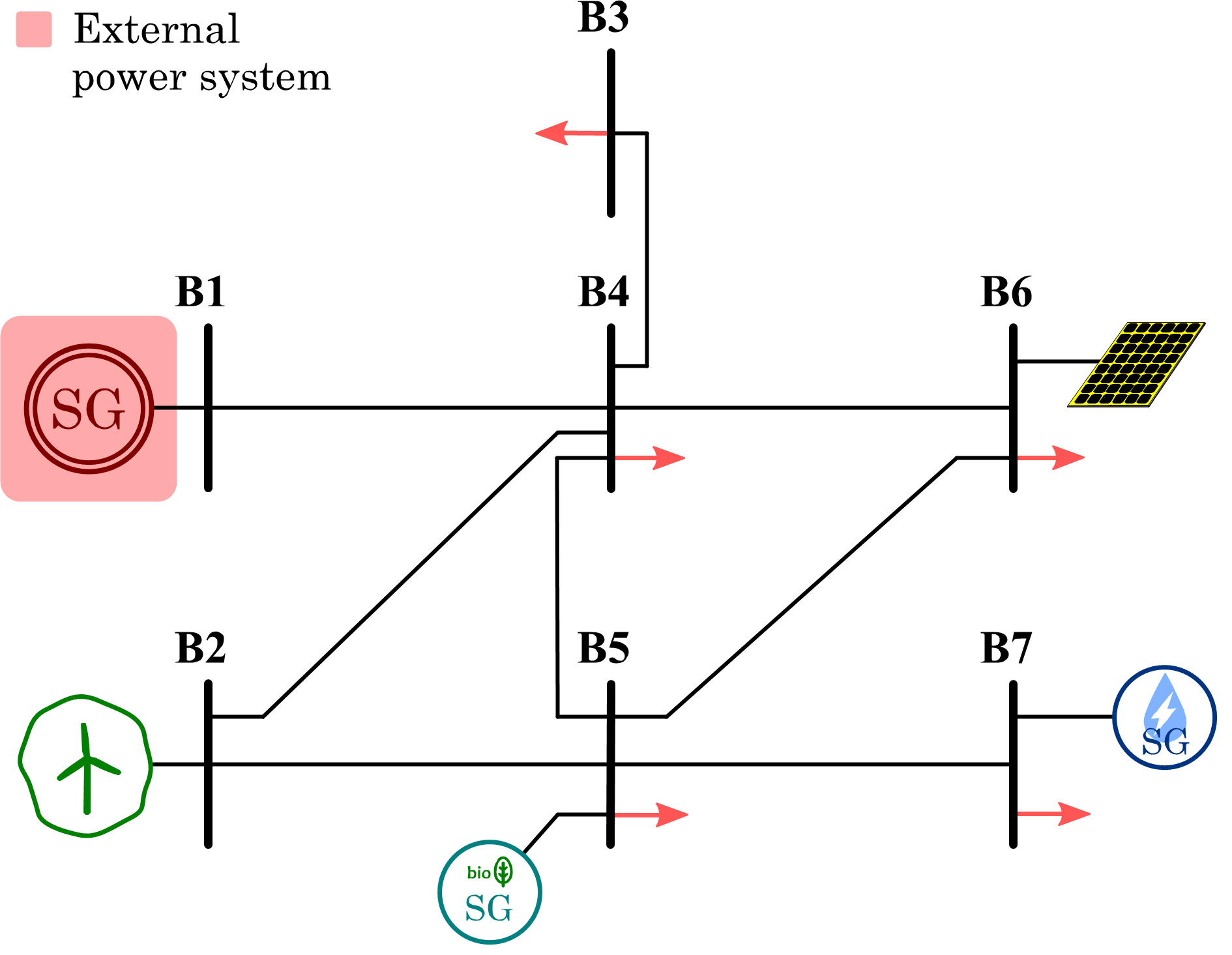}
\caption{Scenario type I}
\label{fig:sce1}
\end{figure}

\begin{figure}
\centering
\includegraphics[scale=0.2]{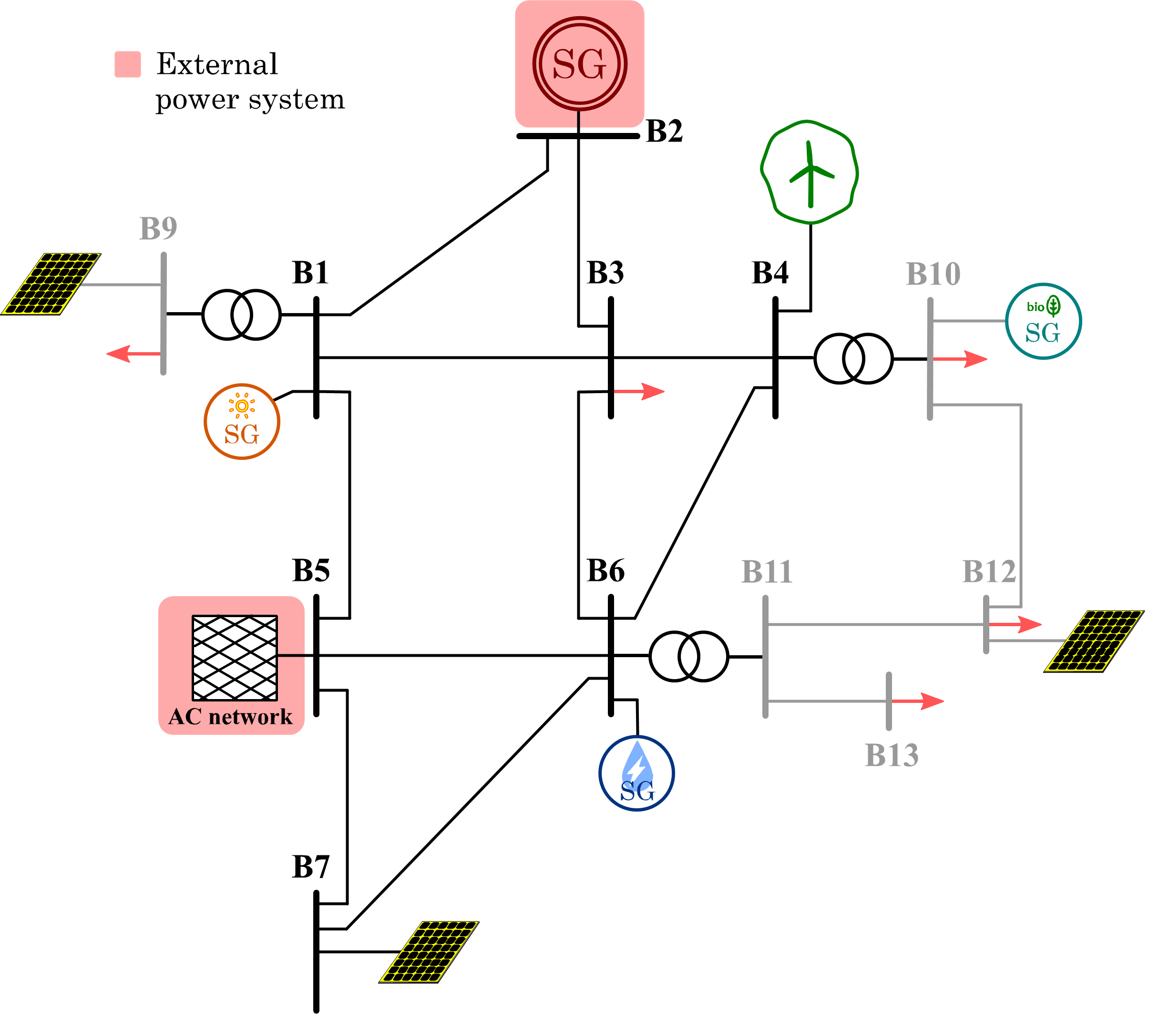}
\caption{Scenarios type II - South Europe}
\label{fig:sce2a}
\end{figure}

\begin{figure}
\centering
\includegraphics[scale=0.2]{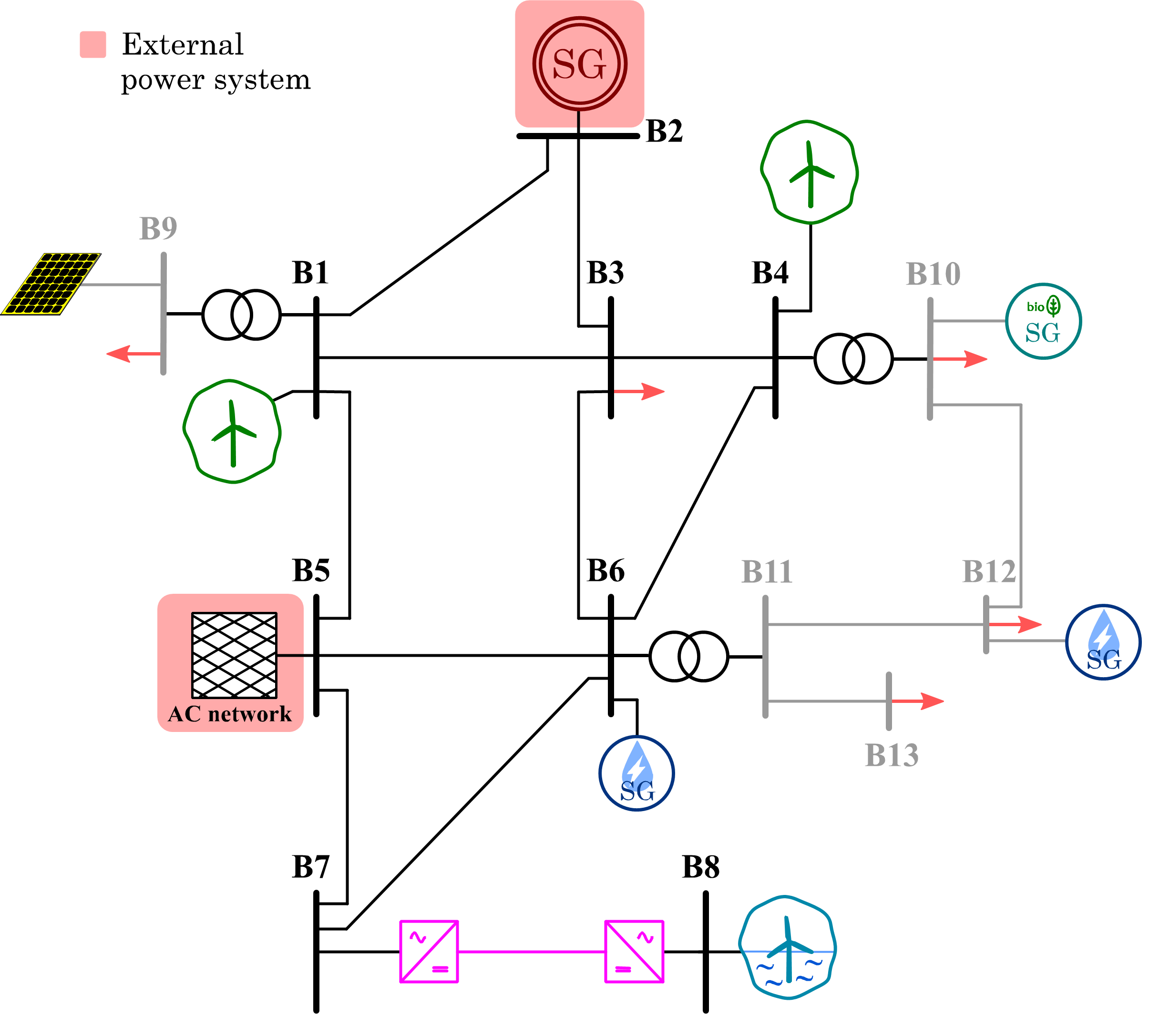}
\caption{Scenarios type II - North Europe}
\label{fig:sce2b}
\end{figure}

\begin{figure}
\centering
\includegraphics[scale=0.2]{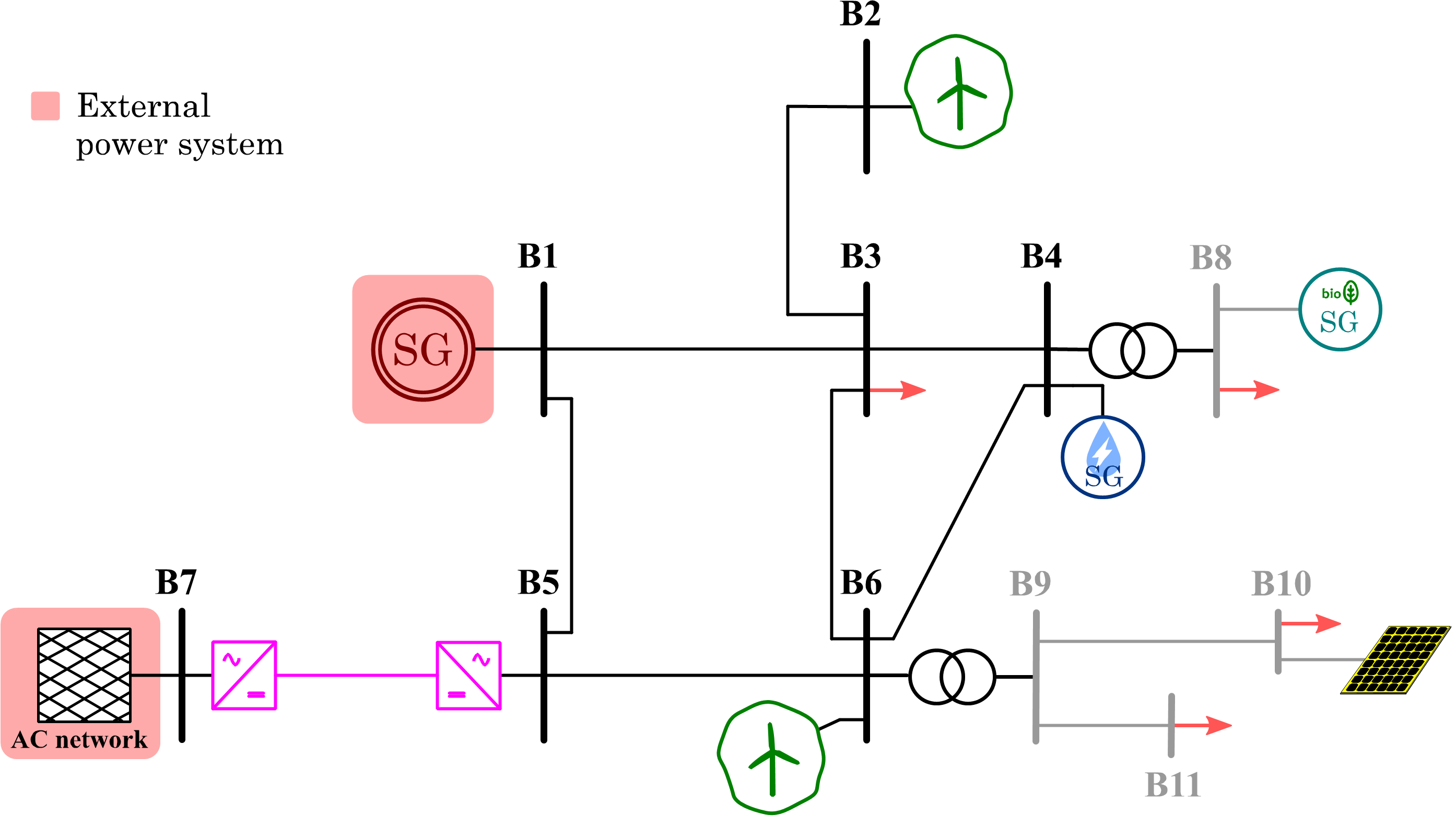}
\caption{Scenarios type III}
\label{fig:sce3}
\end{figure}

The following scenarios have been defined:
\begin{itemize}
\item	Type I: islanded scenarios are in general smaller and simpler as compared to continental scenarios. Therefore, a smaller number of buses (in this case, 7) and a single voltage level is considered for this case (Figure \ref{fig:sce1}).
\item	Type II: the vast majority of scenarios are AC interconnected systems, and they are typically bigger and more meshed. Therefore, a higher number of buses (in this case, 13) and different voltage levels (i.e., transmission and distribution) are considered. Moreover, two different versions of this type of scenario are considered. One corresponds to a typical scenario with good solar resource (for example southern Europe) (Figure \ref{fig:sce2a}), whereas the other corresponds to a typical scenario with good onshore and offshore wind resource (Figure \ref{fig:sce2b}), including HVDC interconnected offshore wind (for example northern Europe).
\item Type III: regarding HVDC interconnected scenarios without AC interconnections, they typically correspond to bigger islands. For that reason, the grid layout considered is slightly more complex, with a higher number of buses as compared to Type I (in this case, 11). Also, different voltage levels are also considered in this case (Figure \ref{fig:sce3}).
\end{itemize}

\subsection{Optimization}
\label{sub:optimization}    

The original concept of VPP germinated from the need of tackling the relatively low competitiveness of the back then emerging non-dispatchable RESs such as wind and solar generation when compared with large, dispatchable conventional generation such as hydro and thermal power plants. Most power system regulators require a minimum bid size to be submitted to electricity market auctions, leaving most stochastic, non-dispatchable RESs out of market, being remunerated by ex-post settlements with generally lower profitability than that of market participants.
To increase this competitiveness is clearly essential to pave the way to an eventual mass integration of such renewable sources. In this vein, aggregation of non-dispatchable RESs  in the form of a VPP as a single offering unit in electricity markets with a total size larger than the minimum bid size becomes apparent. 

An alternative to the VPP concept that is rapidly gaining interest is the installation of electrochemical (battery) energy storage systems (BESS) due to their capability to provide both active and reactive power regulation with very short time responses (down to several tens of milliseconds).  Moreover, the advances in the BESS technology and their modularity imply a remarkable flexibility that allows the installation of BESSs of up to 100 MW and over 100 MWh such as the lithium-ion BESS installed in Hornsdale, Australia, in 2017. The potential of BESSs to provide a large number of ancillary services and to mitigate the impact of the stochastic nature of non-dispatchable RESs, together with the gradual decrease in the price per MW and MWh, justify their current popularity as a solution to increase the competitiveness of non-dispatchable RESs.

However, the main limitations of BESSs that prevent their massive integration in the power grids are their still high installation (capital) costs, their relatively short life span (up to 7 years, or a few thousands of cycles), their intrinsic self-discharge (up to two percentual digits per day of their state of charge), the limited availability of the materials required in their construction, and their negative environmental impact at the time of their disposal.

The RES-based DVPP proposed in this work thus appears as a promising approach to overcome the limitations of BESSs listed above, and as a competitive solution to increase the viability of non-dispatchable RESs.
The DVPP is composed of already installed RESs (both dispatchable and non-dispatchable) and demands that can provide some level of flexibility, thus reducing the installation costs to only the deployment of the communication infrastructure required to coordinate all assets. Moreover, by optimally operating all assets, the increase of non-dispatchable RES competitiveness may even surpass the benefits achieved with BESS solutions.

Electricity generation companies rely mostly on power and energy markets for obtaining revenues from electric energy trading. This is so since the late 1990s, where a transition from optimal control strategies took place worldwide at a remarkable pace \cite{Conejo18}.

Long-term electricity trading is mostly based on bilateral contracts, whereas short-term trading is generally based on competitive auctions or pools. The latter represents over 75\% of the total energy traded \cite{Dagoumas19}.
Short-term markets thus appear as the most relevant for the studies considered in this work given their competitive nature. Short-term electricity trading usually spans a time window of 24 hours, and different pools take place prior to the power delivery. Two main groups of participants are present in such pools, namely generation (electricity producers) and demand (retailers and large consumers). According to the quantity traded, short-term electricity markets can be also categorized into energy markets, in which the market operator gives/receives payments according to the amount of energy supplied to/consumed from the network. Other markets are based on the trading of power for ancillary service provision (mostly active, although reactive power-based markets are starting to become relevant).

The optimal participation of the proposed RES-based DVPP in the aforementioned energy and power markets (operation) will thus be key point in development of the DVPP concept. The impact of uncertainties that characterize the stochastic RESs in the DVPP and market prices will also be duly analyzed. To this aim, robust optimization \cite{Bertsimas04, Baringo16} will be implemented, where uncertainties of stochastic generation and market prices are modeled as confidence bounds and intervals.

Another important aspect is the \textit{market power} of the DVPP. Generally, market offering units based on RESs other than hydroelectric power plants are relatively small. This implies that their auction participation does not change the resulting clearing price of electricity, i.e., they are \textit{price takers}. However, if enough RESs are aggregated in the form of a DVPP, the volume of energy auctioned could be high enough to alter such a price, making the DVPP a \textit{price maker}. In this situation, the DVPP is aware of its own market power, and will naturally tend to alter the price in such a way that its benefits will be maximized. However, the regulatory agent will attempt to minimize the influence of all price makers in the system when clearing the market. This converts the single-stage optimization problem solved for a price taker DVPP into an iterative, multi-level problem when the DVPP is a price maker, with a significant increment in complexity. 

For the aspects mentioned before, it has been assumed that the DVPP configuration was \textit{given}, in the sense that the DVPP is composed of a known set of already installed RESs and flexible demands. However, the determination of which configuration of DVPP (total size, share of the different RES technologies and demands, geographical location of the assets, etc) would maximize the overall profit of the DVPP in electricity markets while minimizing associated risks is the result of an optimization problem that needs to take into account, apart from the concepts outlined above, e.g., the costs associated to the installation and exploitation of the assets. The optimal configuration of the DVPP for the variety of scenarios listed in Section~\ref{sub:scenarios} is also an important output of the DVPP concept.

\section{Dynamic aspects}\label{sectionDynamic}

\subsection{Generator control}
\label{sec:GenoControl}
\begin{figure}
    \centering
    \includegraphics[width=1\linewidth]{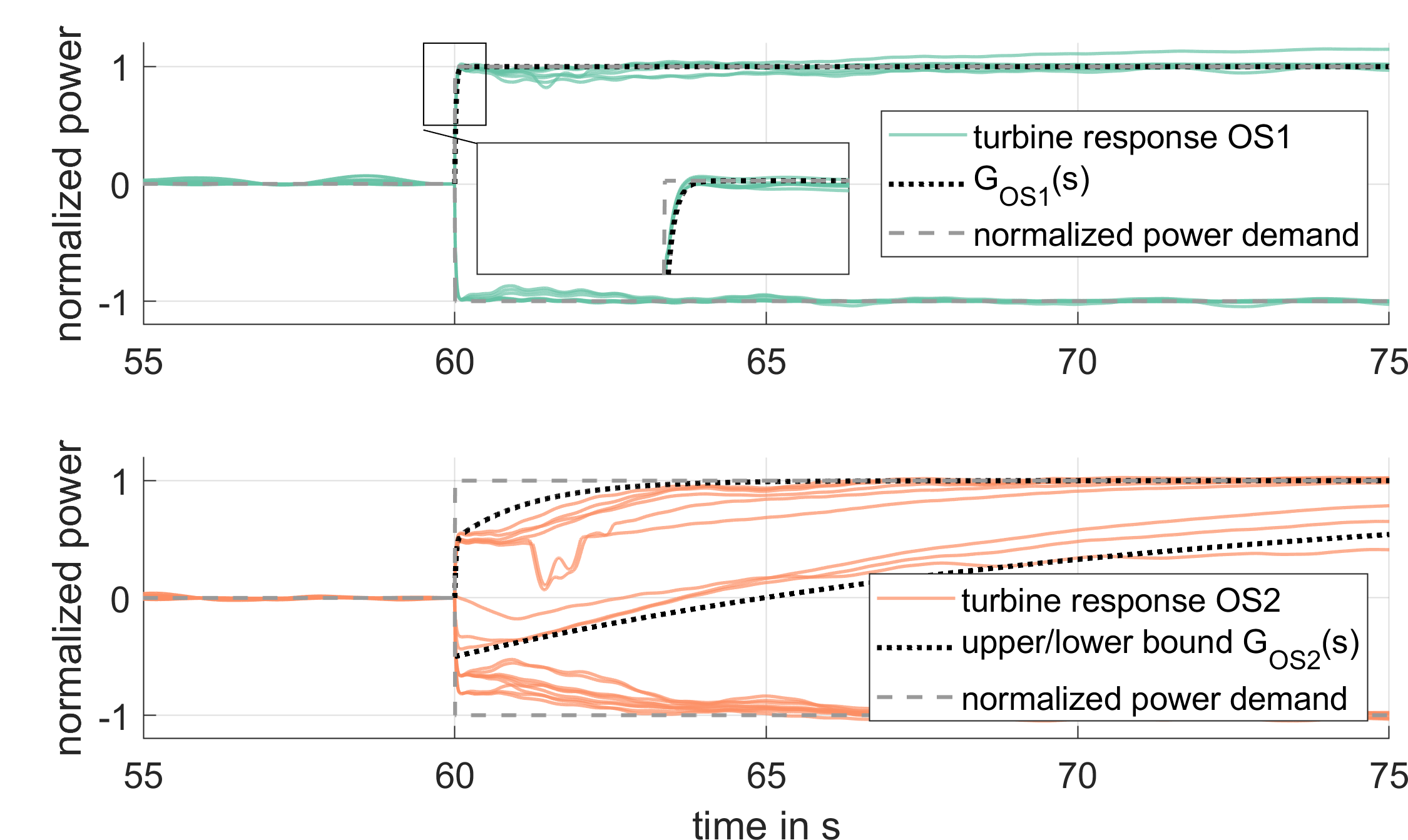}
    \caption{Normalized step response for demand steps of $\Delta P_{ref}$ at constant wind speeds of $v=\{8,\,12,\,16\}$\,m/s}
    \label{fig:WT_power_tracking}
\end{figure}

Dynamics of each RES generator should be managed in order to ensure safe operation (from the material point of view, i.e., keeping currents, voltages and mechanical loads (in case the mechanical structures of wind generators) within technological limits of operation) and contractual obligations. The latter are mainly on the active power production. To ensure that, controls should be implemented to track setpoints for active power, voltage and mechanical speed. As RES are usually connected to the grid by power electronics, supplementary controls for the used power converters and the DC part are needed.\\
In order to be able to operate a RES-based grid control without BESS (see Section~\ref{sub:optimization}), it is necessary that the non-dispatchable power plants (PV, W) also provide a required gradient of active power. The achievable power gradient is thereby strongly dependent on the used control method. For example, the usual control strategy of wind power plants aims for a maximization of the power output in partial-load region and a limitation of power above rated wind speed \cite{Bossanyi2003}. In contrast, compared to the usual strategy, demanded power point tracking leads to an increased operating range, which must be managed by the controller in terms of load reduction with fast response times to be optimized \cite{Poeschke2020}. The operating trajectory that results in a desired power output, however, is not unique and therefore depends on the choice of the operational scheme encoded in the control concept. This can be illustrated by considering the generator power given as
\begin{equation}
	\label{eq:WT_power}
	p(v) = \omega_g(v) \, T_g(v) \, , 
\end{equation}
where $\omega_g$ and $T_g$ are the rotational speed and generator torque, respectively, and $v$ represents the current effective wind speed. From (\ref{eq:WT_power}), it is apparent that a variation of power output to the demand can be achieved by an adjustment of either the rotational speed, the generator torque or both. Consequently, there is a need to study the implications of different operating strategies for power tracking about the structural loading and possible response dynamics in DVPP operation.
To illustrate this, the power demand transmission behavior of a wind turbine generator for two different operating strategies (OS) proposed in \cite{Poeschke2021} will now be briefly presented. In the first concept, termed OS1, the demanded power is achieved by a variation of the generator torque only while keeping the rotational speed at its nominal value at the current wind speed. Contrarily, in OS2 the controller enforces a variation of both, the generator torque and rotational speed to meet the power demand. For comparison of the dynamics, the results are shown in Figure~\ref{fig:WT_power_tracking} as normalized step responses of $\Delta P_{ref}$ by $\Delta P$, where $\Delta P_{ref}$ denotes the demanded and $\Delta P$ the power generated by the wind turbine generator. To assess the dynamics involved, the turbine is faced with instantaneous demand changes while operating in different constant wind conditions at a constant power output of 70 percent of nominal power.  The step wise changes in the power demand $P_{ref}$ are bidirectional, i.e., increase and reduction of the power output demand at steps of \mbox{$\Delta P_{ref} = \{-0.3,\,-0.2,\,-0.1,\,0.1,\,0.2,\,0.3\}$} is conducted. The simulation is repeated for wind speeds of $v=\{8,\,12,\,16\}$\,m/s to roughly cover a range of common operating wind speeds, and in conjunction with the bidirectional steps possibly reveal nonlinear effects. In analogy to the wind turbine control, the PV system should be operated with a demanded power point tracking (DPPT) instead of a maximum power point tracking algorithms (MPPT). \\
With the results obtained, it is thus feasible to use non-dispatchable RES as power generators in the DVPP concept. Thus, based on the results obtained so far, it is feasible to use non-dispatchable RES as power generators in the DVPP concept. Further investigations are being performed specifically for PV power plants with equally realistic scenarios as for the wind turbine generators.
 
\subsection{Grid control}
Ancillary services are important to maintain operation in case of system incident or variation of operating conditions. Till today, few requirements at this level are made for RES. RES from the DVPP should \textit{fully} participate to grid services. This means not only to provide some voltage and frequency help, but for the entire DVPP to be able to participate to actual implemented control schemes in same conditions as the large thermal plants. For this, it is not sufficient to add a supplementary control layer. The control at the generator level should be revised. For example, as mentioned in Section~\ref{sec:GenoControl}, the RES cannot be run with MPPT or DPPT \cite{2} \cite{3} and a reserve should be managed for grid frequency services via \textit{deloading control} (see, e.g., \cite{7}). Moreover, multiplication of the controls needs and actions led us to a \textit{dynamic system view} for the DVPP for modeling, specifications and control levels as shown in the next section. This means a new approach for the control, with a global view of the system and specifications and a maximum coordination in the control actions.

\subsection{Internal redispatch}
\begin{figure}
    \centering
    \includegraphics[scale=0.7]{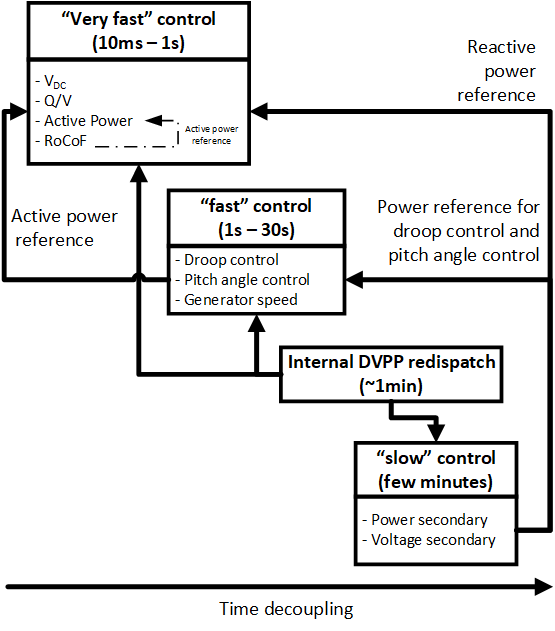}
    \caption{Time decoupling}
    \label{fig:TimeScales}
\end{figure}

Local and grid objectives should be ensured not only for nominal operation but also in case of failure or variation of availability of ressources (sun or wind, for example). The latter variations have an important impact on non-dispatchable RES of the DVPP. To ensure a continuous run of the DVPP (especially for the ancillary services), enough fast \textit{internal redispatch} of the RES resources is needed. This is done via a tool for the \textit{real-time portfolio optimization} of the DVPP. It mainly consists into a security-constraint optimization problem for which both generators and grid constraints (and even overall system constraints) must be taken into account. The time constant of such redispatch loop can be chosen around one minute in order to integrate it in the overall control dynamics given in Fig. \ref{fig:TimeScales} with no parasitic interactions.

\subsection{Interaction with neighbour dynamic elements of the grid}
As mentioned before, DVPP is not necessarily a geographically exclusive grouping. Indeed, between the devices selected to be included in a DVPP, other static and dynamic devices of the grid exist and may have a coupled behaviour with the ones of the DVPP. Also, the same holds for devices in the neighborhood of the geographic perimeter of the DVPP. Such interactions can be globally taken into account in the aggregated control models mentioned in Section \ref{sectionApproaches}. However, certain devices require some particular caution. First, the power electronic based ones may have higher interactions with the converters of the DVPP, especially in case of rapid control of the latter ones (see, e.g., \cite{Milano2018}, \cite{Schoe}). Second, coordinated control with HVDCs and FACTS like SVCs and TCSCs may have a beneficial impact on overall grid performances and avoid oscillations between converters \cite{Munt}. 

\subsection{Dynamics of systems dominated by power electronics}
Modern power systems are increasingly dominated by power electronics, including FACTS devices, HVDC converters, renewable energy interfaces and loads driven by power electronics. The special nature of power electronics compared to synchronous generators have motivated significant research in the last decade. However, the issues related to system dynamics and stability have still to be investigated. The DVPP concept should integrate this in order to be applied to future power systems. 

Novel methodologies will have to be developed both for analysing, simulating and understanding the system under study as well as for the control. 

The development of models of different degrees of complexity and granularity (including non-linear detailed models of power electronics as well as linearized simple models) together with employing co-simulation techniques will allow developing sound analysis and identifying the possibilities and limitations of the considered systems \cite{Prieto2017}, \cite{Prieto2016}. 

As mentioned before, the conventional power system stability definitions, their assessment, and the control thereof through ancillary services rely on the notion of synchrony. The latter is ensured in a conventional grid through large rotational generation and tight control of frequency and voltage predominantly by actuating synchronous machines. The replacement of synchronous machines by power electronics-interfaced generation not only changes the qualitative power system dynamics towards more brittle dynamics (with shorter time scales), but the system is also subject to increasingly many fluctuations (due to variable renewable generation) and with interfaced with ever-more fragile devices (e.g., converter-interfaced generation) that cannot tolerate large fault currents. Finally, also the conventional classification and separation into voltage, angle, and frequency stability does not hold anymore in future power systems that operate possibly far from a nominal equilibrium and are driven by disturbances on all time scales \cite{Milano2018}. 

Novel stability definitions and analysis concepts are required that not only take into account the volatile physics of future power systems but also hard operational limits, accounting e.g. for converter over-currents. Hence, rather than looking for the conventional “stability of a stationary equilibrium point”, one should investigate how disturbances amplify fluctuations in a neighbourhood of the nominal synchronous operation.

Control, including ancillary services, should also be designed in accordance to this new context, as discussed in the next section.

\section{Discussion of approaches}\label{sectionApproaches}

In this section are discussed some main approaches to implement the DVPP concept.

\subsection{Models}
To achieve the system-level view mentioned above, we believe that models should be constructed in a new way. They should be global (include both device and system-level dynamics) but also sufficiently simplified and tractable to cover the DVPP perimeter and to be used for control design. Modern power systems contain hybrid, i.e., both slow and fast dynamics. Fast dynamics comes from power electronics which are systematically used to connect RES to grid and to reinforce transmission grids (with direct-current – HVDC - lines). Modeling should be innovative both for simulation as well as for control purposes. Indeed, to prevent from unnecessary and unmanageable high-dimension of the resulting models, specific approaches to capture dynamics of interest should be used. For simulation, the so-called \textit{co-simulation} method can be used. It mainly consists in simulating a large zone of the power system with different degrees of details for well-chosen sub-zones. For the synthesis of the control models, identification of the models in specific bands of frequency closed linked to the phenomena to be preserved would allow low-order yet highly expressive mathematical models.

The assumptions underpinning these models are also different from the ones used till now in power systems. First, DVPP perimeter is not compliant with classic assumption of split of the grid into very-high and low and medium voltage levels. Indeed, this voltage separation is strong as followed by administrative and control/operation separation since the grids are run by different entities: the high-voltage ones by TSOs and the others by DSOs. Strong hypothesis of non-interaction of controls and separation of data and information are at the basis of this structure. As RES are connected at both voltage levels, to ensure optimality in DVPP definition and run, one should thus envisage DVPP perimeter which include both sides of the grid. Hypothesis mentioned above should be revisited to construct models adequate for this kind of DVPP perimeter.

Next, in perspective to participation to ancillary grid services, a larger view is needed: include future grid dynamics, include not only one DVPP but several to deal with coordination (in secondary controls for voltage and power) and competition in operation, include other dynamic power-electronic based devices like, e.g., HVDC, to deal with interactions and to damp some oscillations which might exist between these devices and the RES generators of DVPP. 

Finally, all aspects should be considered in the DVPP design and control: voltage, frequency, internal robustness/resilience/redispatch (N-1 grid stability). Because of the fast/slow time-scales mentioned above, it is no longer possible (as in classic today approaches) to treat separately these phenomena and the new models should integrate all them.

\subsection{Optimal operation and configuration of DVPP}

The operation of DVPPs is separated into an internal and external operation architectures. DVPPs are currently grouped into technical (TDVPP) and commercial (CDVPP) ones, interacting mutually by allocating power commitment (CDVPP) and rescheduling due to technical reasons (TDVPP).

Although separation between TDVPP and CDVPP has been made, the nowadays massively available amount of information and data has not been systematically considered yet. Further, most of the models used to formulate the internal set-points (i.e., references for the dynamic control loops such as active or reactive power set-points) for the economic operation are non-linear models, leading to mixed integer non-linear optimization problems. These problems can, for the case of power systems, be however transformed into mixed integer linear problems for which efficient and scalable (possibly suboptimal) approaches exist by now \cite{Morales2014}, \cite{Morales2013}, \cite{Tejada2018}, \cite{Cerisola2009}.

Large-scale deployment of DVPPs further complicates computation of centralized set-points. Aggregation and disaggregation strategies will become of high importance to improve applicability. Combination of centralized and distributed internal set-points might be an interesting alternative. Large-scale deployment also heavily increases burdens due to additional system and grid constraints. Particularly, dynamic constraints have not been addressed widely.
Finally, large-scale deployment turns DVPPs from price taker to price maker, requiring not only models of the DVPP but also of the complete power system.
Large-scale deployment of DVPPs with dispatchable and non-dispatchable RES requires new modeling and solution approaches. The mutual impact of DVPPs and the power system will be addressed in an iterative manner. A simplified power system model will be used, where generation is grouped according to its generation technology and by making use of standardized techno-economic parameters for each group. The concept of clustered unit commitment can be used \cite{Morales2019}.

Uncertainties of stochastic renewable sources and of electricity market prices need to be duly taken into account in the optimization problem in order to obtain accurate information of expected revenues and costs for a given DVPP operation. Uncertainties will be taken into account through the so-called \textit{robust optimization} \cite{Bertsimas04, Baringo16}, where the objective of the optimization problem is the maximization of the revenue (minimization of the costs) for the worst-case realization of the uncertain profiles.

Modelling will include a linear representation of the network by using a DC power flow model for active power \cite{IIT}. Dynamic constraints will be included as well. Models will be linearized through equivalent, linear expressions. Internal and external management algorithms (i.e., internal dispatch and external market participation) will be addressed by proposing appropriate aggregation and disaggregation strategies, and by combining centralized and distributed controls. Finally, reactive power constraints will also be considered by modelling the network with a linearized AC optimal power flow model.

\subsection{Advanced model-based DVPP control}

The traditional approach to automatic control is based on a first-principle model of the physical system to be controlled. These models come in different parameterizations, e.g., frequency-domain transfer functions, state-space models, or higher-order differential equations. Uncertainties are inevitable in the modeling process, and are handled in model-based design through either robust or adaptive control techniques \cite{Zhou}, \cite{Bemp}, \cite{Wu}. Both the literature on power systems modeling as well as model-based control are mature and highly developed. 

However, sometimes first-principle physical models are too complex to be useful for control design (e.g., the wake interactions inside a wind farm), too high-dimensional and large-scale for the considered control objective (e.g., to damp inter-area oscillations it is not required to know the detailed continental power grid model), or simply too cumbersome to calibrate (e.g., precisely fitting the values of all passives inside a converter). In such cases one would opt for reduced-order models (e.g., an area equivalent), identify non-physical (e.g., an ARMA model for load behaviour) models from time series data, or directly go for a data-driven control design leveraging recent advances and methodologies from the machine learning community. Another way to construct a \textit{control-model} for the DVPP is to extract dynamics of interest of the overall system into an as simple as possible (from the state dimensional point of view) mathematical object \cite{Mari2021}.

From the control point of view, there are two ways to tackle DVPP control problem. First, with a \textit{centralized control} which takes a maximum benefit from the unified \textit{system-view} of the DVPP and the surrounding power system. A control model which takes into account all dynamics and interactions is used along with classic robust control methods and, as a consequence, one can expect maximum coordination, performances and robustness. The price to pay is the use of several measurements (and some of them could be from distant generators) and a lack of resilience in case of failure of one actuator (RES generator). Next, as opposite philosophy of control, \textit{decentralized approaches} can be used. They use only local measurements to design control loops around each device. Among these extreme cases that will be presented below, intermediate control solutions can be envisaged like, for example, centralized synthesis of controllers with decentralized implementation.

\subsubsection{Centralized and coordinated control}

\begin{figure}
    \centering
    \includegraphics[scale=0.5]{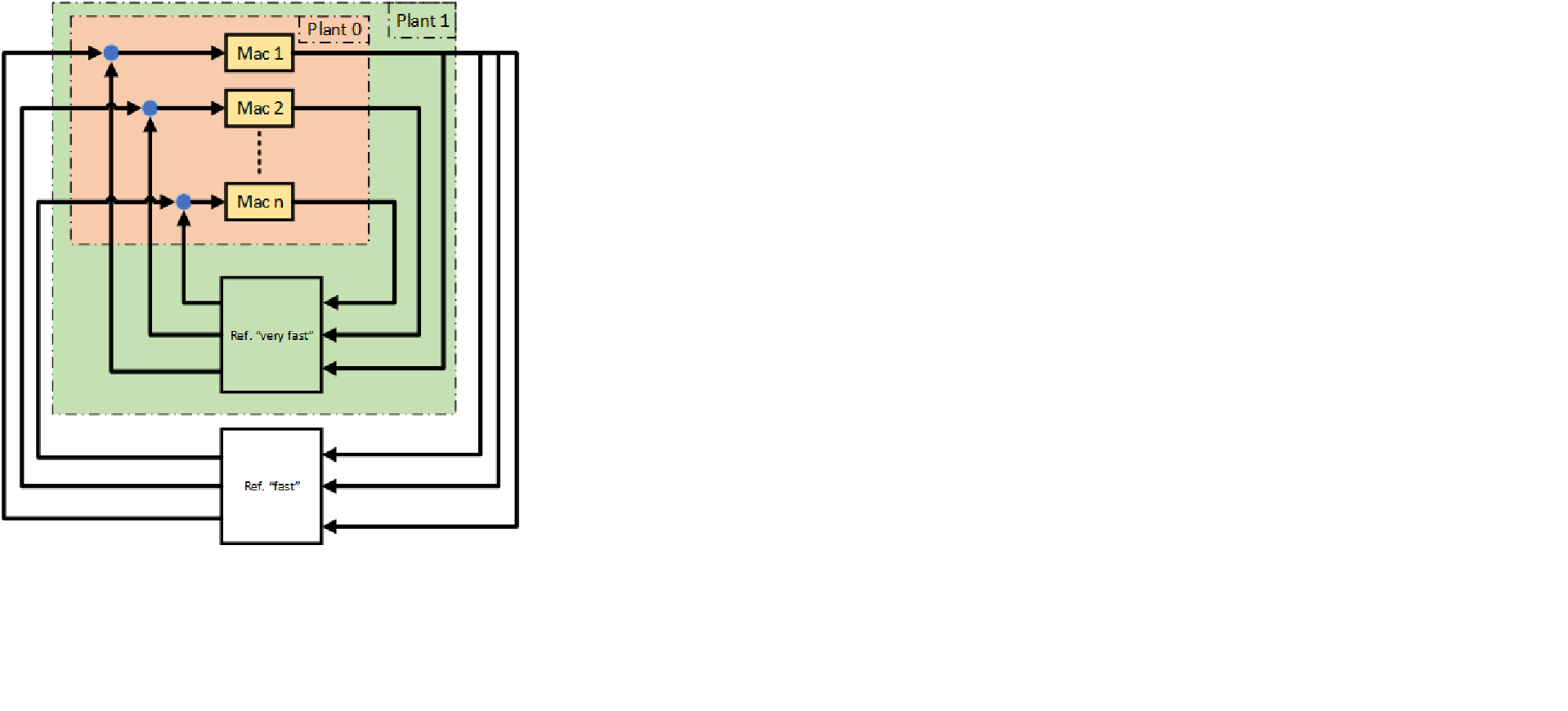}
    \caption{Centralized control}
    \label{fig:CentralizedControl}
\end{figure}

Control schematized in Fig. \ref{fig:CentralizedControl} is a centralized approach to handle several RES generators according to the time-decoupling of the dynamics of the phenomena shown in Fig. \ref{fig:TimeScales}. Compared with the classic vector control, the control is not structured around each actuator, but according to the time response (frequency band) of the actuators and open-loop plant dynamics \cite{Mari2021}. 

Several stages of control are proposed according to the time scales. The closed-loop obtained at one stage is the plant for the next stage. In this way a hierarchical and sequential synthesis is possible, with, at each level, account for the faster controls of lower levels and with minimal risk of parasitic dynamic interactions. Notice also that this strategy is compliant with actual organization of controls in power systems (structured in primary/secondary layers) and opens the way to direct integration of RES into existing power systems controls and market mechanisms.

\subsubsection{Decentralized control}

\begin{figure}
    \centering
     \includegraphics[scale=0.5]{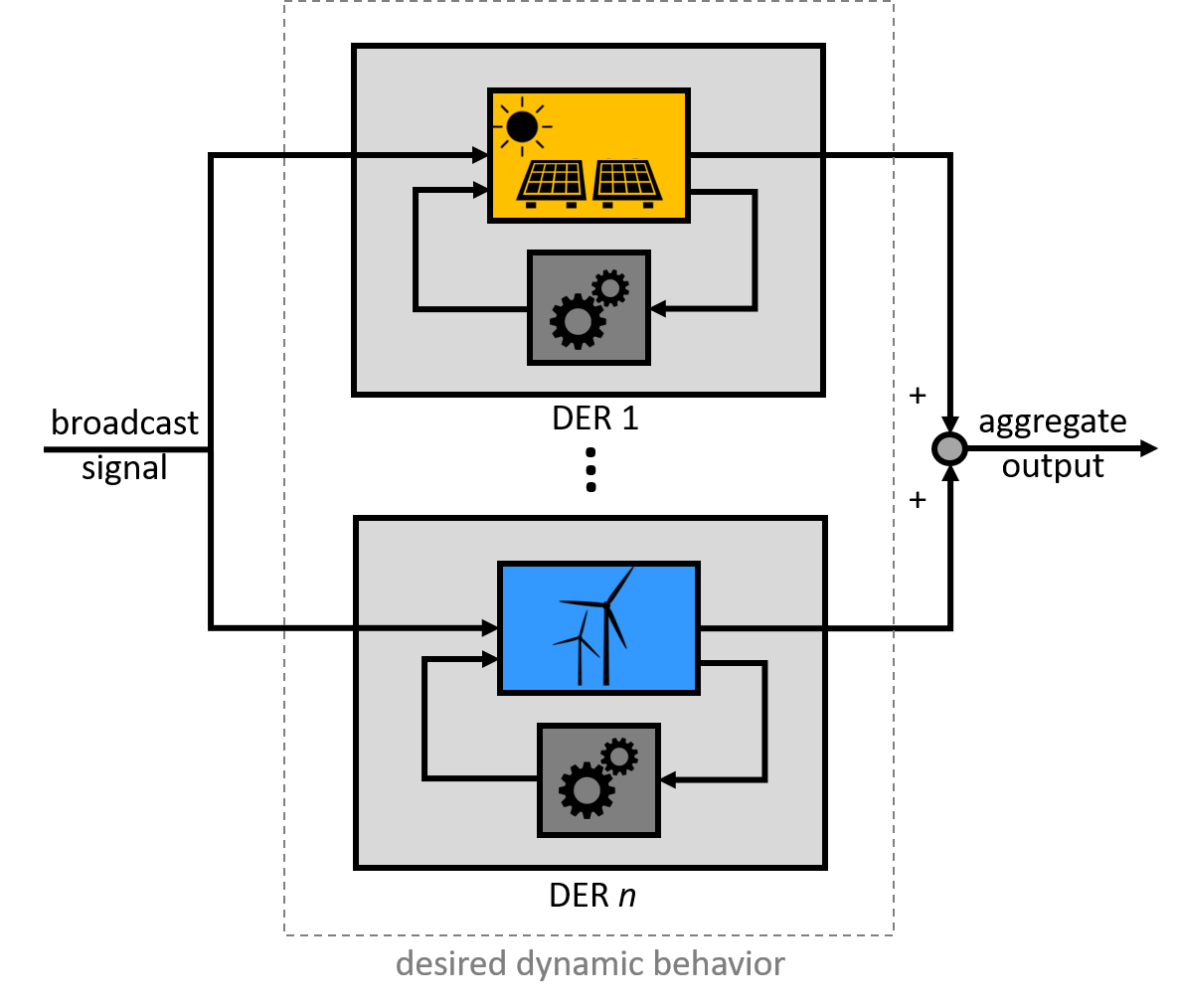}
    \caption{Decentralized control}
    \label{fig:DecentralizedControl}
\end{figure}

In parallel to the centralized control approach, we also pursue a fully decentralized  approach, where the device-level controllers employ only local measurements as well as a few selected global broadcast signals. As a prototypical example, consider a fast-frequency response provided by a DVPP in grid-following mode; see Fig. \ref{fig:DecentralizedControl} for an illustration. In this scenario, a selected global frequency measurement signal is broadcast to the local controllers of the devices comprising the DVPP. The local controllers are designed so that the aggregate response of the DVPP -- from the frequency measurement to the aggregate power output -- meets the aggregate dynamic specification, in this case a desired droop and virtual inertial response. Aside from the aggregate design specification, the local controllers also have to take the local device-level limitations into account, such as bandwidth limitations, energy and power constraints, over-current limitations, and so on. In summary, the DVPP control design is posed as a {\em decentralized matching problem}: local controllers should meet the aggregate specification subject to device-level constraints.

We pursue two distinct approaches to this problem. Our first approach is based on a divide-and-conquer strategy: the aggregate specification is disaggregated to purely local specifications by means of dynamic participation factors (a dynamic extension of the well-known static participation factors). Next local matching controllers are tasked to meet the disaggregated specification; see \cite{Bjork2021} for a preliminary exploration of this approach and a case study coordinating hydro and wind power for fast frequency response. Our second approach falls square in between the centralized and decentralized control paradigms: optimal and structured controllers are designed in a centralized fashion but so that they allow for a decentralized implementation. For both approaches we envision also grid-forming as well as adaptive strategies that adapt online to changing  conditions inside the DVPP, e.g., fluctuations of wind and solar production.

\section{Conclusions}\label{sectionConclusions}

The DVPP concept presented in this paper will pave the way to top-down solutions to increase RES penetration (dispatchable and non-dispatchable) in actual and future power systems. It allows one to treat all the aspects -static, dynamic, optimality/efficiency - at once and in a coordinated way in order to provide TSOs, DSOs and generators with knowledge, models and tools. Controls for both local (production) and grid (ancillary services) objectives specifications will be provided for the actual power systems in full compliance with the existing regulations schemes, as well as for future power systems with massive RES penetration and low inertia systems. 

More specifically, the outputs will be:
\begin{itemize}
	\item optimality criteria to define the perimeter/portfolio of DVPP both for long term and real-time application
	\item new controllers to allow RES to fully contribute to ancillary services ultimately enabling system stability
	\item new business cases for the optimal operation and configuration of DVPP
	\item regulatory recommendations to enable DVPP development and operation in conjunction with the generators run in a classic way
	\item assessment of economic competitiveness of the DVPP compared with solutions combining variable RES with electrochemical storage
	\item new stability definitions and methodologies for stability analysis and assessment
\end{itemize}

Part of the approaches mentioned in Section \ref{sectionApproaches} are innovatory and will be fully reported in forthcoming publications. Other are based on classic methods which are used in a new manner in the specific DVPP context.


\begin{thebibliography}{1}

\bibitem{EUroadmap} European Comission, “Energy roadmap 2050,” Luxembourg: Publications Office of the European Union, 2012.

\bibitem{VPP3} C. Huang, D. Yue, J. Xie, Y. Li, and K. Wang, 
Economic Dispatch of Power Systems with Virtual Power Plant Based Interval Optimization Method,  CSEE JOURNAL OF POWER AND ENERGY SYSTEMS, VOL. 2, NO. 1, MARCH 2016.

\bibitem{VPP4} D. Koraki and K. Strunz, Wind and Solar Power Integration in Electricity Markets and Distribution Networks Through Service-Centric Virtual Power Plants, IEEE TRANSACTIONS ON POWER SYSTEMS, VOL. 33, NO. 1, JANUARY 2018.

\bibitem{VPP5} N. Pourghaderi, M. Fotuhi-Firuzabad, M. Moeini-Aghtaie, and M. Kabirifar, Commercial Demand Response Programs in Bidding of a Technical Virtual Power Plant, IEEE TRANSACTIONS ON INDUSTRIAL INFORMATICS, VOL. 14, NO. 11, NOVEMBER 2018.

\bibitem{VPP6} S. Camal, A. Michiorri and G. Kariniotakis, Optimal Offer of Automatic Frequency Restoration Reserve From a Combined PV/Wind Virtual Power Plant, IEEE TRANSACTIONS ON POWER SYSTEMS, VOL. 33, NO. 6, NOVEMBER 2018.

\bibitem{VPP_Spain1} E.L. Miguelez, I.E. Cortes, L. Rouco, G.L. Camino, «  An overview of ancillary services in Spain », Electric Power Systems Research, Vol.  78, pp. 515–523, 2008.

\bibitem{VPP_Spain2} I. Egido, F. Fernández-Bernal, L. Rouco, Member,  « The Spanish AGC System: Description and Analysis », IEEE TRANSACTIONS ON POWER SYSTEMS, VOL. 24, NO. 1, FEBRUARY 2009.

\bibitem{Milano2018} F. Milano, F. Dörfler, G. Hug, D.J. Hill, G. Verbic, Foundations and Challenges of Low-Inertia Systems, Proc. of PSCC 2018.

\bibitem{Kouki2018} M. Kouki, B. Marinescu, F. Xavier, Eigencalculation of Coupling Modes in Large-Scale Interconnected Power Systems with High Power Electronics Penetration, Proc. PSCC 2018.

\bibitem{Arioua2016} L. Arioua, B. Marinescu, « Robust grid‐oriented control of high voltage DC links embedded in an AC transmission system », International Journal of Robust and Nonlinear Control 26 (9), pp. 1944-1961, 2016.

\bibitem{Munt2017} I. Munteanu, B. Marinescu, F. Xavier, « Analysis of the interactions between close HVDC links inserted in an AC grid », IEEE General Meeting, 2017.

\bibitem{Bel2018} M. Belhocine, B. Marinescu, F. Xavier, “A Generic Model for Power Park Modules for Both Transient and Small-signal Stability Analysis, Proc. PSCC 2018.

\bibitem{Conejo18} Conejo, A. J.,  Baringo, L. Power System Operations. Cham, Switzerland: Springer International Publishing, 2018.

\bibitem{Dagoumas19} A, Dagoumas, "Impact of Bilateral Contracts on Wholesale Electricity Markets: In a Case Where a Market Participant Has Dominant Position". \textit{Applied Sciences}, 9(3), 382, 2019.

\bibitem{Bertsimas04} D. Bertsimas, M. Sim, "The Price of Robustness", \textit{Operations Research},  52 (1) 35-53, 2004.

\bibitem{Baringo16} M. Rahimiyan, L. Baringo, "Strategic Bidding for a Virtual Power Plant in the Day-Ahead and Real-Time Markets: A Price-Taker Robust Optimization Approach," in \textit{IEEE Transactions on Power Systems}, vol. 31, no. 4, pp. 2676-2687, July 2016.

\bibitem{Bossanyi2003}
E.~A. Bossanyi, "Wind turbine control for load reduction", in \textit{Wind Energy}, vol. 6, pp. 229–244, 2003.

\bibitem{Poeschke2020} F. P{\"o}schke, E. Gauterin, M. K{\"{u}}hn, J. Fortmann, H. Schulte, Load mitigation and power tracking capability for wind turbines using linear matrix inequality-based control design, in \textit{Wind Energy}, vol. 23, pp 1792--1809, 2020.

\bibitem{Poeschke2021}
F. P{\"o}schke, H. Schulte, Evaluation of different APC operating concepts considering turbine loading and response time for grid support, in \textit{Wind Energy Science Conference (WESC)}, Hannover, Germany, May 2021,.  

\bibitem{2}
Fernández-Guillamón, A., Gómez-Lázaro, E., Muljadi, E., Molina-García, Á., Power systems with high renewable energy sources: A review of inertia and frequency control strategies over time. Renewable and Sustainable Energy Reviews, 115, 109369, 2019.

\bibitem{3}
Hohm, D. P., Ropp, M. E., Comparative study of maximum power point tracking algorithms. Progress in photovoltaics: Research and Applications, 11(1), 47-62, 2003.

\bibitem{7}
Dreidy, M., Mokhlis, H., Mekhilef, S., Inertia response and frequency control techniques for renewable energy sources: A review. Renewable and sustainable energy reviews, 69, 144-155, 2017.

\bibitem{Schoe} 
K. Schoenleber, E. Prieto, S. Rates O. Gomis, Extended Current Limitation for Unbalanced Faults in MMC–HVDC–connected Wind Power Plants, IEEE Transactions on Power Delivery, Vol 33, Num 4, pp 1875-1884, 2018.

\bibitem{Munt}
I. Munteanu, B. Marinescu, F. Xavier, « Analysis of the interactions between close HVDC links inserted in an AC grid », IEEE General Meeting, 2017.

\bibitem{Prieto2017} 
E. Prieto, A. Junyent, G. Clariana, O. Gomis, Control Design of Modular Multilevel Converters in Normal and AC Fault Conditions for HVDC grids, Electric Power Systems Research, Vol 152, Num 11, pag 424-437, 2017.

\bibitem{Prieto2016}
E. Prieto, A. Egea, S. Frekiasl , O. Gomis, DC Voltage Droop Control Design for Multiterminal HVDC Systems Considering AC and DC Grid Dynamics, IEEE Transactions on Power Delivery, Vol 31, Num , pag 575 - 585, 2016.

\bibitem{Morales2014}
G. Morales-España, A. Ramos, J. García-González. An MIP formulation for joint market-clearing of energy and reserves based on ramp scheduling. IEEE Transactions on Power Systems. vol. 29, no. 1, pp. 476-488, January 2014.
 
\bibitem{Morales2013}
G. Morales-España, J.M. Latorre, A. Ramos. Tight and compact MILP formulation for the thermal unit commitment problem. IEEE Transactions on Power Systems. vol. 28, no. 4, pp. 4897-4908, November 2013. 

\bibitem{Tejada2018}
D.A. Tejada, P. Sánchez, A. Ramos. Security constrained unit commitment using line outage distribution factors. IEEE Transactions on Power Systems. vol. 33, no. 1, pp. 329-337, January 2018. 

\bibitem{Cerisola2009}
S. Cerisola, Á. Baíllo, J.M. Fernández-López, A. Ramos, R. Gollmer. Stochastic power generation unit commitment in electricity markets: a novel formulation and comparison of solution methods. Operations Research. vol. 57, no. 1, pp. 32-46, January 2009

\bibitem{Morales2019} 
G. Morales-Espana and D. A. Tejada-Arango, "Modelling the Hidden Flexibility of Clustered Unit Commitment," in IEEE Transactions on Power Systems, Volume: 34, Issue: 4, pp.  3294 - 3296, July 2019.

\bibitem{IIT}
F.M. Echavarren, E. Lobato, L. Rouco, Full active-reactive DC power flow model with loss compensation, Working paper IIT IIT-16-119A, July 2016.

\bibitem{Zhou}
Zhou, K., Doyle, J. C., Glover, K., Robust and optimal control (V. 40, p. 146). NJ: Prentice hall, 1996.

\bibitem{Bemp}
Bemporad, A.,  Morari, M., Robust model predictive control: A survey. In Robustness in identification and control(pp. 207-226). Springer, London, 1999.

\bibitem{Wu}
Wu, X., Dörfler, F.,  Jovanović, M. R., Input-output analysis and decentralized optimal control of inter-area oscillations in power systems. IEEE Transactions on Power Systems, 31(3), 2434-2444, 2016.

\bibitem{Mari2021}
B. Marinescu, E. Kamal and H-T. Ngo, A System Model-Based Approach for the Control of Power Park Modules for Grid Voltage and Frequency Services, Preprint http://arxiv.org/abs/2107.02000, 2021.

\bibitem{Bjork2021}
Björk, Joakim, Karl Henrik Johansson, and Florian Dörfler. "Dynamic Virtual Power Plant Design for Fast Frequency Reserves: Coordinating Hydro and Wind." arXiv preprint arXiv:2107.03087, 2021).
\end{thebibliography}

\end{document}